\documentclass[aps,prl,twocolumn,superscriptaddress,showpacs,preprintnumbers,amsmath,amssymb]{revtex4}
\usepackage{graphicx} 
\usepackage{dcolumn}  

\graphicspath{{ps}}

\renewcommand{\arraystretch}{1.1}

\usepackage{rotating}

\begin{document}

\preprint{\vbox{ \hbox{   }
            \hbox{Belle Preprint 2010-22}
            \hbox{KEK Preprint 2010-37}
}}


\title{ \quad\\[1.0cm]
{\boldmath {\large Measurement of the decay $B^0\to\pi^-\ell^+\nu$ and determination of $|V_{ub}|$}}
}

\affiliation{Budker Institute of Nuclear Physics, Novosibirsk}
\affiliation{Faculty of Mathematics and Physics, Charles University, Prague}
\affiliation{University of Cincinnati, Cincinnati, Ohio 45221}
\affiliation{Justus-Liebig-Universit\"at Gie\ss{}en, Gie\ss{}en}
\affiliation{The Graduate University for Advanced Studies, Hayama}
\affiliation{Hanyang University, Seoul}
\affiliation{University of Hawaii, Honolulu, Hawaii 96822}
\affiliation{High Energy Accelerator Research Organization (KEK), Tsukuba}
\affiliation{Hiroshima Institute of Technology, Hiroshima}
\affiliation{University of Illinois at Urbana-Champaign, Urbana, Illinois 61801}
\affiliation{Indian Institute of Technology Guwahati, Guwahati}
\affiliation{Institute of High Energy Physics, Chinese Academy of Sciences, Beijing}
\affiliation{Institute of High Energy Physics, Vienna}
\affiliation{Institute of High Energy Physics, Protvino}
\affiliation{Institute for Theoretical and Experimental Physics, Moscow}
\affiliation{J. Stefan Institute, Ljubljana}
\affiliation{Kanagawa University, Yokohama}
\affiliation{Institut f\"ur Experimentelle Kernphysik, Karlsruher Institut f\"ur Technologie, Karlsruhe}
\affiliation{Korea Institute of Science and Technology Information, Daejeon}
\affiliation{Korea University, Seoul}
\affiliation{Kyungpook National University, Taegu}
\affiliation{\'Ecole Polytechnique F\'ed\'erale de Lausanne (EPFL), Lausanne}
\affiliation{Faculty of Mathematics and Physics, University of Ljubljana, Ljubljana}
\affiliation{University of Maribor, Maribor}
\affiliation{Max-Planck-Institut f\"ur Physik, M\"unchen}
\affiliation{University of Melbourne, School of Physics, Victoria 3010}
\affiliation{Nagoya University, Nagoya}
\affiliation{Nara Women's University, Nara}
\affiliation{National Central University, Chung-li}
\affiliation{National United University, Miao Li}
\affiliation{Department of Physics, National Taiwan University, Taipei}
\affiliation{H. Niewodniczanski Institute of Nuclear Physics, Krakow}
\affiliation{Nippon Dental University, Niigata}
\affiliation{Niigata University, Niigata}
\affiliation{University of Nova Gorica, Nova Gorica}
\affiliation{Novosibirsk State University, Novosibirsk}
\affiliation{Osaka City University, Osaka}
\affiliation{Panjab University, Chandigarh}
\affiliation{University of Science and Technology of China, Hefei}
\affiliation{Seoul National University, Seoul}
\affiliation{Sungkyunkwan University, Suwon}
\affiliation{School of Physics, University of Sydney, NSW 2006}
\affiliation{Tata Institute of Fundamental Research, Mumbai}
\affiliation{Excellence Cluster Universe, Technische Universit\"at M\"unchen, Garching}
\affiliation{Toho University, Funabashi}
\affiliation{Tohoku Gakuin University, Tagajo}
\affiliation{Tohoku University, Sendai}
\affiliation{Department of Physics, University of Tokyo, Tokyo}
\affiliation{Tokyo Metropolitan University, Tokyo}
\affiliation{Tokyo University of Agriculture and Technology, Tokyo}
\affiliation{CNP, Virginia Polytechnic Institute and State University, Blacksburg, Virginia 24061}
\affiliation{Wayne State University, Detroit, Michigan 48202}
\affiliation{Yonsei University, Seoul}
  \author{H.~Ha}\affiliation{Korea University, Seoul} 
  \author{E.~Won}\affiliation{Korea University, Seoul} 
  \author{I.~Adachi}\affiliation{High Energy Accelerator Research Organization (KEK), Tsukuba} 
  \author{H.~Aihara}\affiliation{Department of Physics, University of Tokyo, Tokyo} 
  \author{T.~Aziz}\affiliation{Tata Institute of Fundamental Research, Mumbai} 
  \author{A.~M.~Bakich}\affiliation{School of Physics, University of Sydney, NSW 2006} 
  \author{V.~Balagura}\affiliation{Institute for Theoretical and Experimental Physics, Moscow} 
  \author{E.~Barberio}\affiliation{University of Melbourne, School of Physics, Victoria 3010} 
  \author{A.~Bay}\affiliation{\'Ecole Polytechnique F\'ed\'erale de Lausanne (EPFL), Lausanne} 
  \author{K.~Belous}\affiliation{Institute of High Energy Physics, Protvino} 
% ...
\collaboration{The Belle Collaboration}

\begin{abstract}
We present a measurement of the charmless semileptonic decay
$B^0\to\pi^-\ell^+\nu$ 
using a data sample containing 657$\times 10^6$ $B\bar{B}$ events 
collected with the Belle detector at the KEKB asymmetric-energy
$e^+e^-$~collider operating near the
$\Upsilon(4S)$~resonance. We determine the total branching fraction of the decay, 
$\mathcal{B}(B^0\to\pi^-\ell^+\nu)=(1.49\pm 
0.04{(\mathrm{stat})}\pm 0.07{(\mathrm{syst})})\times 10^{-4}$. We also report 
a new precise measurement of the differential decay rate, 
and extract the Cabibbo-Kobayashi-Maskawa matrix element $|V_{ub}|$ 
using model-independent and model-dependent approaches. 
From a simultaneous fit 
to the measured differential decay rate and lattice QCD results, 
we obtain $|V_{ub}|=(3.43\pm 0.33)\times 10^{-3}$, 
where the error includes both experimental and theoretical uncertainties.
\end{abstract}


\maketitle
\tighten
{\renewcommand{\thefootnote}{\fnsymbol{footnote}}}
\setcounter{footnote}{0}

Weak transitions among quark flavors in the standard model (SM) are
described by the Cabibbo-Kobayashi-Maskawa (CKM)
matrix~\cite{Kobayashi:1973fv}, in which $|V_{ub}|$ is one of the least
known elements. Precise measurements of the values of the CKM matrix 
elements are necessary 
to probe the quark mixing mechanism of the SM and to search for
possible physics beyond the SM. The magnitude of the CKM element $V_{ub}$ can be
determined from exclusive $b\to u \ell \nu$ semileptonic decays, of which 
$B^0\to\pi^-\ell^+\nu$~\cite{CC} yields the most precise value 
for $|V_{ub}|$. The differential rate of this decay can be expressed in terms of
$|V_{ub}|$ and the form factor $f_+(q^2)$, where $q^2$ is the square 
of the momentum transferred from the $B$ meson to the outgoing leptons, 
$q^2=(p_\ell+p_\nu)^2$~\cite{Neubert:1993mb}. 
The present theoretical understanding of $f_+(q^2)$ is limited, which is 
a significant source for systematic uncertainty in the extraction 
of $|V_{ub}|$ from this decay. 
Predictions have been obtained in unquenched lattice QCD
~\cite{Dalgic:2006dt,Okamoto:2004xg}, 
in light cone sum rule (LCSR) theory~\cite{Ball:2004ye} 
and in relativistic quark models~\cite{Scora:1995ty}. 
However, these predictions typically assume a specific shape for $f_+(q^2)$ 
and provide reliable predictions only in a limited $q^2$ range 
(lattice QCD is valid near $q^2$ maximum, 
while LCSR is reliable near the minimum value of $q^2$). 
Recently, it has been shown that 
a determination of $|V_{ub}|$ independent of a form factor shape calculation can 
be achieved by simultaneously fitting the measured $q^2$ spectrum and lattice QCD results 
computed near the zero recoil of $q^2$ range~\cite{Boyd:1994tt;Boyd-Savage:1997,Bailey:2008wp}, 
resulting in $|V_{ub}|=(3.38\pm 0.36)\times 10^{-3}$ using the experimental data 
in Ref.~\cite{Aubert:2006px} for the decay $B^0\to\pi^-\ell^+\nu$ 
where the error includes both theoretical and experimental uncertainties. 
The experimental uncertainty is a 6\% while the theoretical contribution 
is estimated to be an 8.5\%~\cite{Bailey:2008wp}. 
In addition, Ref.~\cite{Aubert:2010px} reports $|V_{ub}|=(2.95\pm 0.31)\times 10^{-3}$ 
by combining measurements of $B^0\to\pi^-\ell^+\nu$ and $B^+\to\pi^0\ell^+\nu$; 
here the error contains a 3\% contribution from the branching fraction measurement, 
a 5\% from the shape of the $q^2$ spectrum measured in data, 
and an 8.5\% from the theoretical normalization. 
Here we describe a study of the decay $B^0\to\pi^-\ell^+\nu$ 
and measure the branching fraction and the $q^2$ spectrum. 
We then compare with other recent studies of this
decay~\cite{Aubert:2006px,Aubert:2010px,Athar:2003yg,Hokuue:2006nr,Aubert:2005cd,Aubert:2006ry}. 
The differential branching fraction is measured in 13 bins of $q^2$, 
and $|V_{ub}|$ is determined using both model-independent 
and model-dependent approaches.

The Belle detector~\cite{:2000cg,Kurokawa:2001nw} is a large-solid-angle magnetic spectrometer that
consists of a silicon vertex detector (SVD), a 50-layer central drift
chamber (CDC), an array of aerogel threshold Cherenkov counters (ACC),
a barrel-like arrangement of time-of-flight scintillation counters
(TOF), and an electromagnetic calorimeter composed of CsI(Tl)
crystals (ECL) located inside a superconducting solenoid coil that
provides a 1.5~T magnetic field.  An iron flux-return located outside of
the coil is instrumented with resistive plate chambers 
to detect $K_L^0$ mesons and to identify muons (KLM).

The data sample corresponds to an integrated luminosity of 605~fb$^{-1}$ 
taken at a center-of-mass (c.m.) energy near the
$\Upsilon(4S)$~resonance, containing 657$\times 10^6$ $B\bar B$~pairs. 
For the first sample of 152$\times 10^6$ $B\bar B$~events, an inner detector
configuration with a 2.0 cm beampipe and a 3-layer SVD was used, 
while a 1.5 cm beampipe, a 4-layer SVD and a small-cell inner drift chamber were used 
to record the remaining
505$\times 10^6$ $B\bar B$ pairs~\cite{Natkaniec:2006rv}. Another
68~fb$^{-1}$ data sample taken at a c.m. energy 60~MeV below the resonance is used to study 
the continuum background, $e^+e^-\to q\bar q$, where $q=u,d,s,c$. 
Monte Carlo (MC)~\cite{evtgen,geant3} simulated events equivalent 
to at least ten times the integrated luminosity were generated to model the signal.
Samples equivalent to ten times and six times the integrated luminosity were generated 
to simulate the two largest background components, 
$b \rightarrow c$ decays and continuum, respectively. 
To simulate rare $b \rightarrow u$ decays, samples equivalent to twenty times 
the integrated luminosity were generated. 
Final state radiation (FSR) from charged particles in the final state 
is modeled using the PHOTOS package~\cite{Barberio:1993qi}.

The decay $B^0\to\pi^-\ell^+\nu$ is reconstructed from 
pairs of oppositely charged leptons and pions. Electron candidates are
identified using the ratio of the energy detected in the ECL to the
track momentum, the ECL shower shape, position matching between the track 
and ECL cluster, the energy loss in the CDC, and the response of the
ACC counters~\cite{Hanagaki:2001fz}. 
Bremsstrahlung photons emitted close to the electron direction are 
reconstructed and used to correct the electron momentum~\cite{Schwanda:2006nf}. 
Muons are identified based on
their penetration range and transverse scattering in the KLM
detector~\cite{Abashian:2002bd}. In the momentum region relevant to
this analysis, charged leptons are identified with an efficiency of
about 90\% while the probability to misidentify a pion as an electron
(muon) is 0.25\% (1.4\%). 
Pion candidates are selected with an efficiency of 85\% 
and a kaon misidentification probability of 19\%, 
based on the responses of the CDC, ACC and TOF subdetectors. 
All charged particles are required to originate from the
interaction point (IP) and to have associated hits in the SVD. The pion
and lepton candidates are fitted to a common vertex and the confidence
level of the fit is required to be greater than 1.0\%. 
The electron (muon) is required to have a laboratory frame momentum greater
than 0.8 GeV/$c$ (1.1 GeV/$c$). 

The missing energy and momentum in the c.m. frame are
defined as $E_\mathrm{miss}\equiv 2E_\mathrm{beam}-\sum_i E_i$ and
$\vec p_\mathrm{miss}\equiv-\sum_i\vec p_i$, respectively, where
$E_\mathrm{beam}$ is the beam energy in the c.m. frame, 
and the sums include all charged and neutral particle candidates in the event. 
A threshold energy of 50 (100) MeV is required for photon candidates in the 
central (side) region of the ECL. 
The neutrino 4-momentum is taken to
be $p_{\nu}=(|\vec p_\mathrm{miss}|,~\vec p_\mathrm{miss})$, 
since the determination of $\vec p_\mathrm{miss}$ 
is more accurate than that of the missing energy. 
To select events compatible with the signal decay mode, 
we require $|Q_\mathrm{total}|\leq 3$, 
where $Q_\mathrm{total}$ is the net charge of the event, 
and $E_\mathrm{miss}>0$ GeV. 
We denote the combined system of the signal pion and lepton as $Y$. 
The kinematics of the decay constrain the cosine of the angle between 
the $B$ and $Y$ directions in the c.m.~frame, defined by
$\cos\theta_{BY}=({2E_\mathrm{beam}E_Y-m_{B}^2-M_Y^2})/(2|\vec p_B||\vec p_Y|)$,
where $m_{B}$ and $|p_B| =\sqrt{E_\mathrm{beam}^2 - m_B^2}$ refer to the mass
and momentum of the $B$ meson, and $E_Y$, $M_Y$, and $p_Y$ refer to the
energy, mass, and momentum of the reconstructed $Y$. 
Background, on the other hand, is not similarly constrained. 
In what follows we require $|\cos\theta_{BY}| \le 1$.
Signal candidates are classified by their beam-energy-constrained mass, 
$M_\mathrm{bc}=\sqrt{E^2_\mathrm{beam}-|\vec p_\pi+\vec p_\ell+\vec p_\nu|^2}$, 
and energy difference, 
$\Delta E = E_\mathrm{beam}-(E_\pi+E_\ell+E_\nu)$. 
Candidates outside of the signal region, defined by the requirements 
$M_\mathrm{bc}>5.19$~GeV/$c^2$ and $|\Delta E|<1$~GeV, are rejected. 
To suppress background from the continuum, 
the ratio of second to zeroth Fox-Wolfram
moments~\cite{Fox:1978vu} is required to be less than 0.35. Background
from $J/\psi\to\mu^+\mu^-$ decays with one muon misidentified as a pion is
rejected by vetoing events with a $Y$~mass between 
3.07~GeV/$c^2$ and 3.13~GeV/$c^2$.
\begin{figure}[b]
\includegraphics[width=0.253\textwidth]{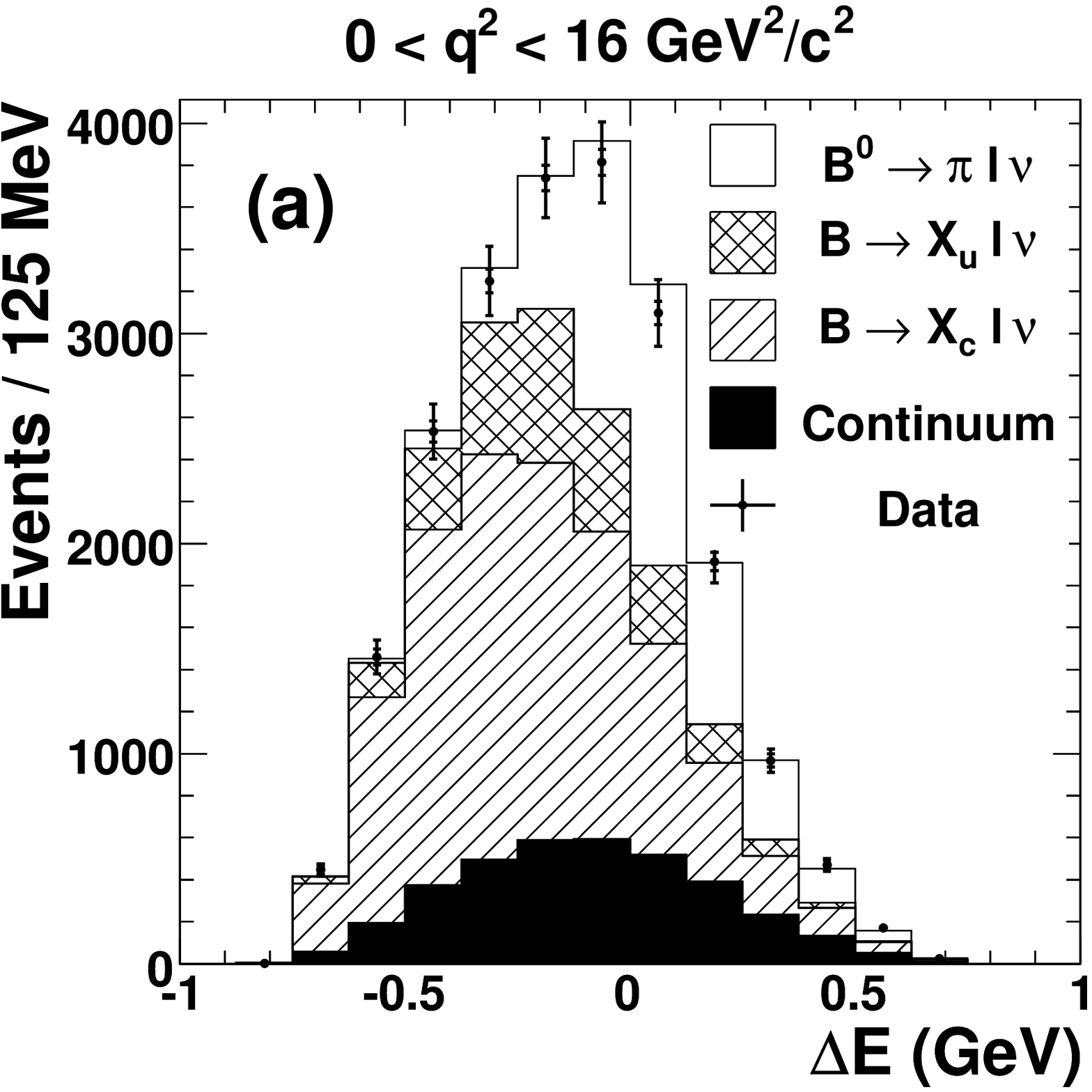}
\hspace{-6.5mm}
\includegraphics[width=0.253\textwidth]{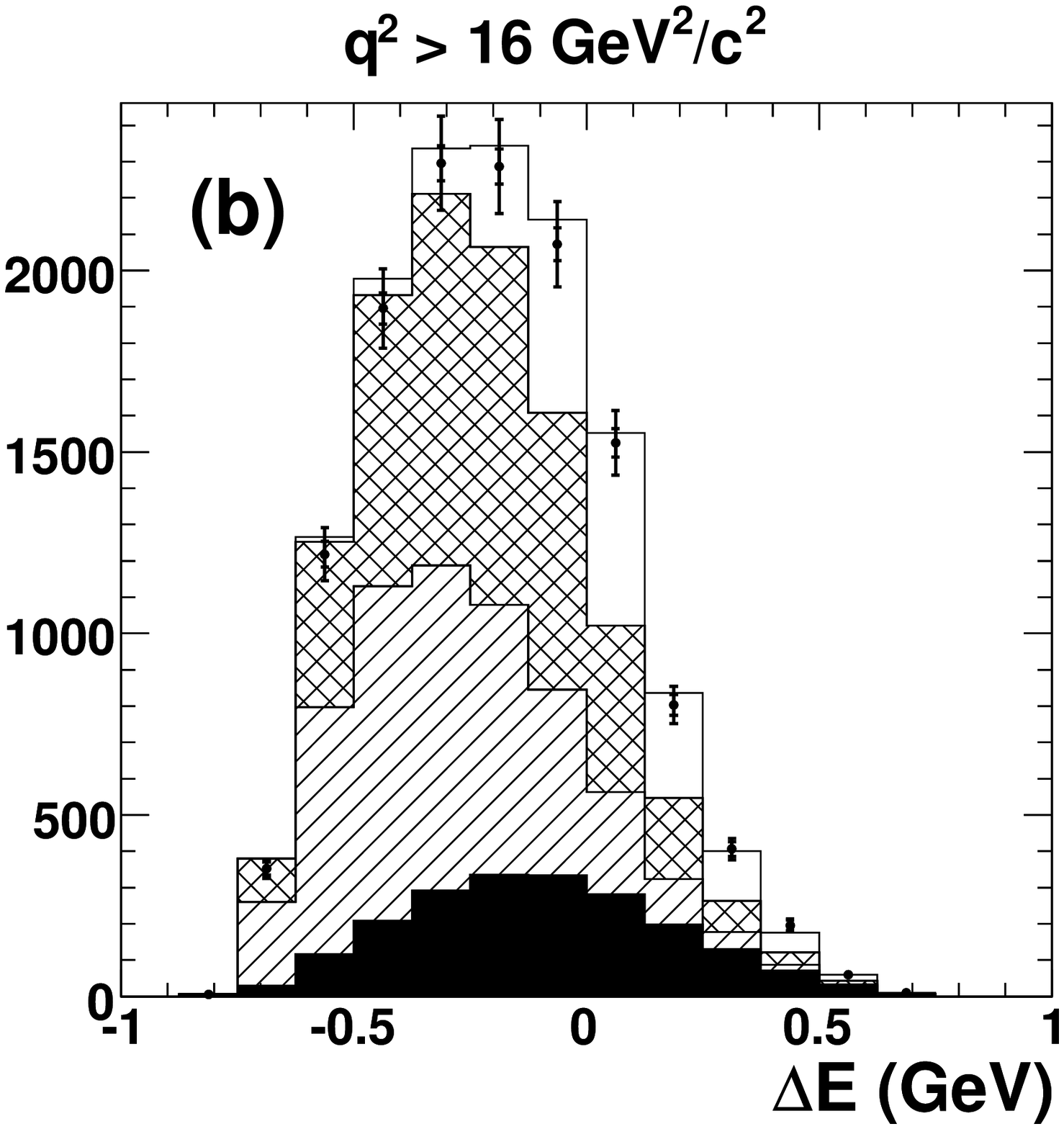} \\
\vspace{-1.7mm}
\includegraphics[width=0.253\textwidth]{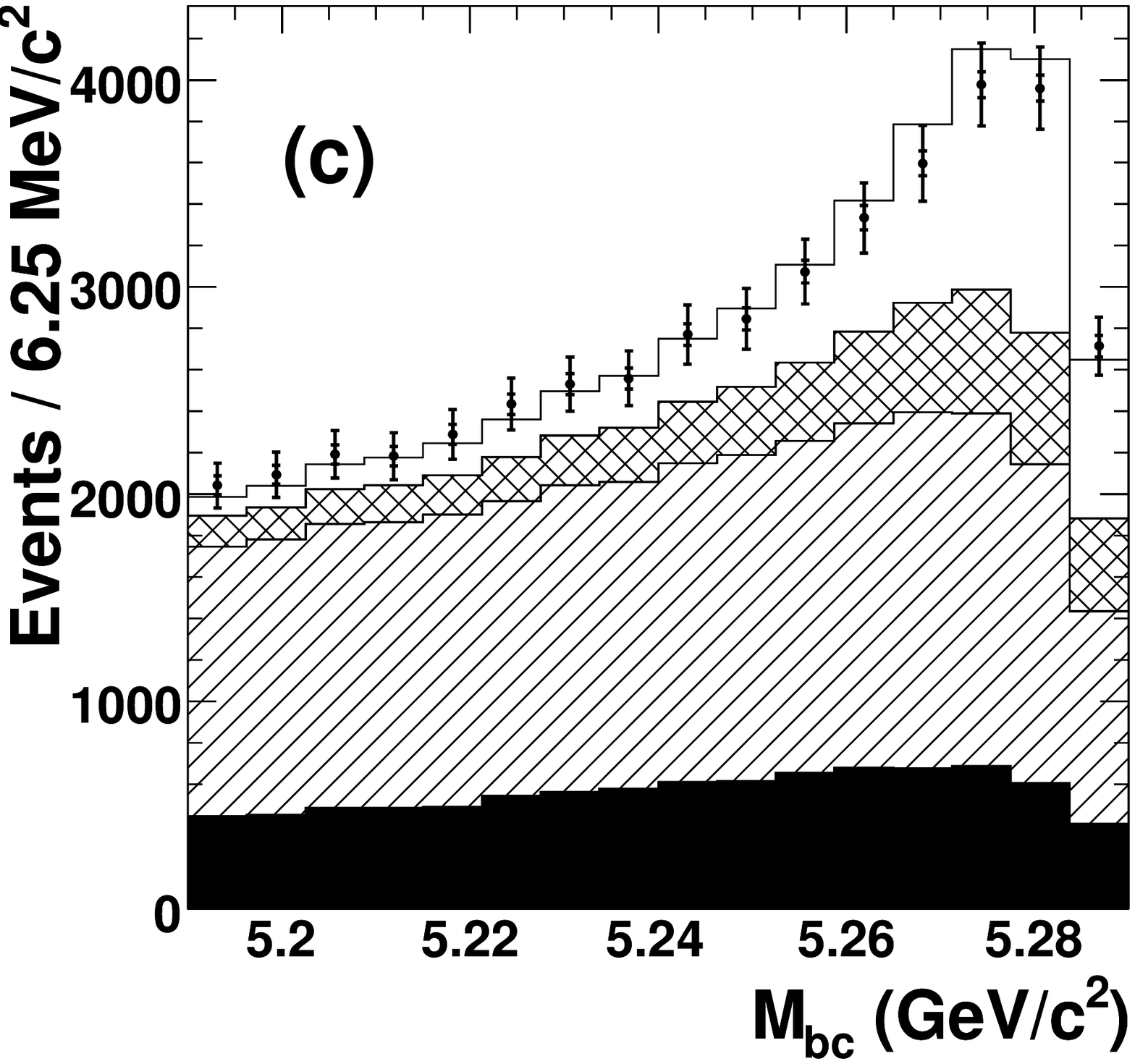}
\hspace{-6.5mm}
\includegraphics[width=0.253\textwidth]{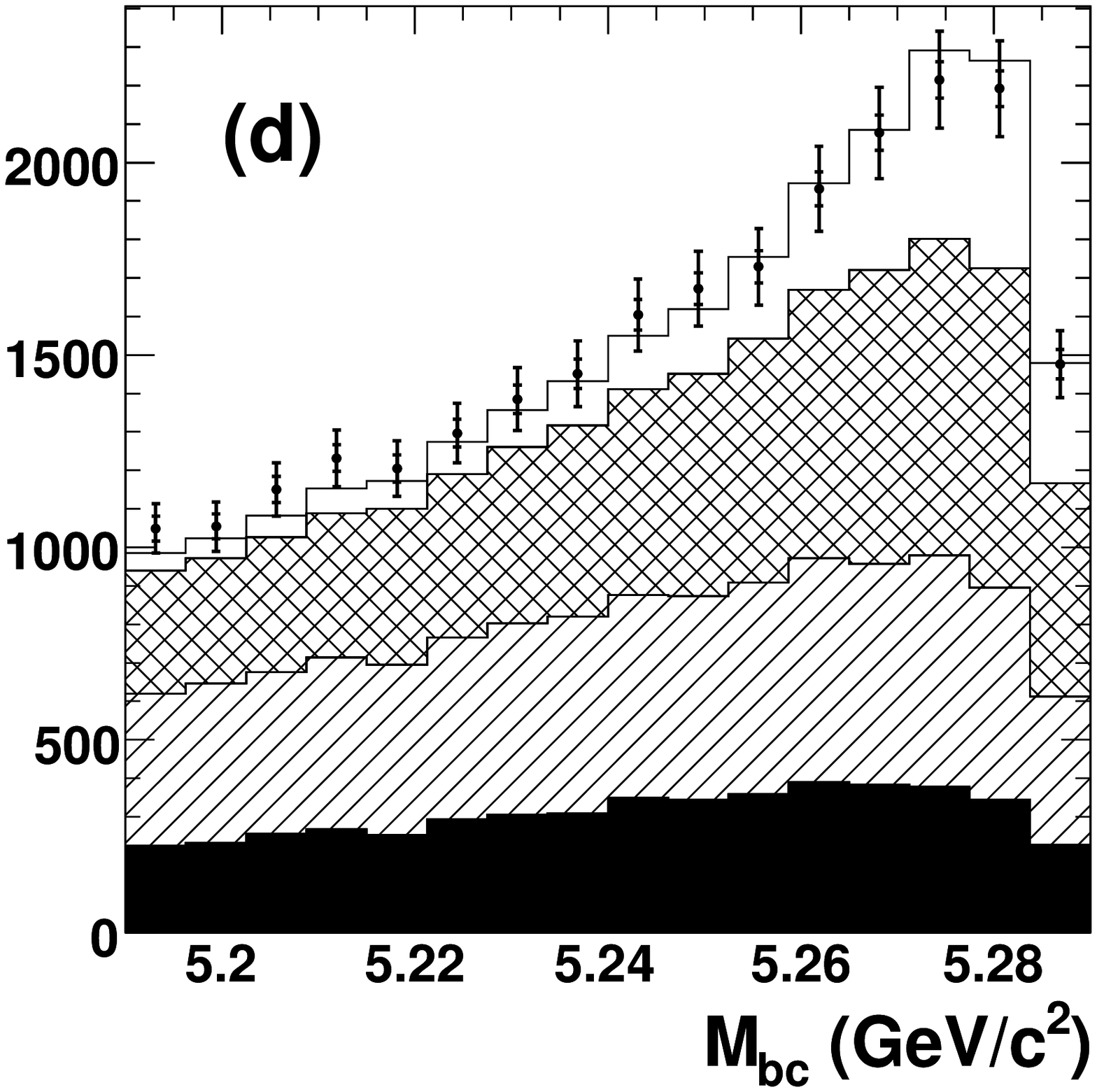}
\vspace{-5mm}
\caption{Fit projections (a,b) in $\Delta E$ with
  $M_\mathrm{bc}>5.27$~GeV/$c^2$, and (c, d) in $M_\mathrm{bc}$ with
  $|\Delta E|<0.125$ GeV. The projections (a,c) and (b,d) show the
  regions $q^2<16$~GeV$^2/c^2$ and $q^2>16$~GeV$^2/c^2$,
  respectively. The points with error bars are $\Upsilon(4S)$~data,
  the histograms are (from top to bottom)
  $B^0\to\pi^-\ell^+\nu$~signal (open), $B\to X_u\ell\nu$ (cross-hatched), $B\to X_c\ell\nu$ (hatched) and continuum background (black-filled).
  The smaller error bars are statistical only while the larger ones include systematic 
  uncertainties.} 
\label{fig1}
\end{figure}
The sample of signal candidates is divided into 13 bins of $q^2$ 
from 0 to 26.4 GeV$^2/c^2$ (the bin width is 2 GeV$^2/c^2$, except for the
last bin). 
The value of $q^2$ is calculated as the square of the difference
between the 4-momenta of the $B$ meson and that of the pion. 
As the $B$~direction is 
only kinematically constrained to lie on a cone around the $Y$~direction,
we take a weighted average over four different possible configurations of the 
$B$~direction~\cite{Aubert:2006cx}. 
Background is further suppressed
by applying selection criteria as a function of $q^2$ to the following quantities: the angle between
the thrust axis of the $Y$~system and the thrust axis of the rest of
the event; the angle of the missing momentum with respect to the beam
axis; the helicity angle of the $\ell \nu$~system~\cite{Cote:2004ch}; 
and the missing mass squared of the event,
$M^2_\mathrm{miss}=E^2_\mathrm{miss}-\vec p^{~2}_\mathrm{miss}$. 
The helicity angle is the angle between the lepton direction 
and the direction opposite to the $B$ meson 
in the  $\ell \nu$ rest frame. 
These selections are optimized separately in each bin of $q^2$ by maximizing 
the figure-of-merit $S/\sqrt{(S+B)}$, where $S$ ($B$) is the expected number
of signal (background) events. 

The fraction of events that have multiple candidates is 66\%. 
To remove multiple signal candidates in a single event, 
the candidate with the smallest $\ell \nu$ helicity angle
is selected. 
After imposing all selections described above, 
the reconstruction efficiency for signal ranges from 7.7\% to 15.0\% 
over the entire $q^2$ range.
The fraction of the self-cross-feed component, 
in which one or more of the signal 
tracks are not correctly reconstructed, is 3.5\%.


The signal yield is determined by performing a two-dimensional, 
binned maximum likelihood fit to the $(M_\mathrm{bc},\Delta E)$~plane
in 13 bins of $q^2$~\cite{Barlow:1993dm}. 
Background contributions
from $b\to u\ell\nu$, $b\to c\ell\nu$ and non-$B\bar B$~continuum are
considered in the fit. 
Probability density functions (PDFs) corresponding to these fit 
components are obtained from MC simulations. 
To reduce the number of free parameters, 
the $q^2$ bins of the background components are grouped into coarser bins: 
four bins for $b\to u\ell\nu$, and three bins for $b\to c\ell\nu$. 
The choice of the binning was chosen from the total statistical error, 
number of parameters to fit, and the complexity of the fits. 
The $q^2$~distribution of the continuum MC~\cite{pythia} simulation 
is reweighted to match the corresponding distribution 
in off-resonance data. 
For this procedure, a continuum MC sample about 60 times the integrated luminosity 
of the off-resonance data is used. 
The continuum normalization is fixed to the scaled number of 
off-resonance events, 52928 events. 
Including signal yields in each $q^2$ bin, 
there are 20 free parameters in the fit.
\renewcommand{\arraystretch}{1.3}
\begin{table*}[t]
\caption{Values of $\Delta\mathcal{B}(q^2)$ and relative uncertainties
  (\%). The uncertainties in MC input parameters are given separately for
	branching fractions (BF) and form factors (FF).}
\begin{ruledtabular}
\begin{tabular}
{lccccccc}
$q^2$(GeV$^2/c^2$)                     & 0 - 6 & 6 - 12 & 12 - 18 & 18 - 26.4 & 0 - 16 & 16 - 26.4 &  Total \\
\hline
$\Delta \mathcal{B}$ ($\times$ 10$^7$)   & 391.19 & 434.25 & 389.47 & 279.18 & 1096.34 &  397.75 & 1494.09 \\
\hline
Detector effects               & 3.4 & 3.5 & 3.5 & 3.5   & 3.4  & 3.5    & 3.4  \\
Physics parameters (BF)        & 0.8 & 0.7 & 0.6 & 0.7   & 0.6  & 0.6    & 0.6  \\
Physics parameters (FF)        & 1.9 & 1.7 & 1.9 & 1.8   & 1.3  & 1.8    & 1.1  \\
Continuum correction           & 4.4 & 2.3 & 3.4 & 2.3   & 2.1  & 2.6    & 1.8  \\
Other sources                  & 2.1 & 2.5 & 2.4 & 2.4   & 2.1  & 2.3    & 2.0  \\
\hline
Total statistical error          & 5.3 & 3.9 & 4.8 & 6.1   & 3.0  & 5.3    & 2.6  \\
\hline
Total error                    & 8.2 & 6.5 & 7.5 & 8.1   & 5.7  & 7.5    & 5.2  \\
\end{tabular}
\end{ruledtabular}
\label{sys1}
\end{table*}

We obtain $21486\pm 548$ signal events, 
$52543\pm 1148$ $b\to u\ell\nu$ events, 
and $161829\pm 976$ $b\to c\ell\nu$ background events. 
These yields agree well with the expectations from MC simulation studies. 
The $\chi^2/\mathrm{n.d.f.}$ of the fit is
$2962/3308$. The projections of the fit result in $\Delta E$ and
$M_\mathrm{bc}$ are shown in Fig.~\ref{fig1} for the regions
$q^2<16$~GeV$^2/c^2$ and $q^2>16$~GeV$^2/c^2$. 
Bin-to-bin migrations
due to $q^2$~resolution are corrected by applying the
inverse detector response matrix~\cite{UNFOLD} to the measured partial
yields. The partial branching fractions~$\Delta\mathcal{B}$ are
calculated using the signal efficiencies obtained from MC simulation. The
total branching fraction~$\mathcal{B}$ is the sum of partial branching fractions
taking into account correlations when calculating the errors. 
We find
$\mathcal{B}(B^0\to\pi^-\ell^+\nu)=(1.49\pm 0.04{(\mathrm{stat})}\pm 0.07{(\mathrm{syst})})\times 10^{-4}$, 
where the first error is statistical and the second error is systematic. 
This result is significantly more precise than our previous measurement~\cite{Hokuue:2006nr} 
with $B\to D^{(*)}\ell^+\nu$ tags on a 253 fb$^{-1}$ data sample.

To estimate the systematic uncertainties on $\Delta\mathcal{B}$, 
we include the following contributions: the uncertainties in
lepton and pion identification, the charged particle reconstruction, 
the photon detection efficiency, 
and the requirement on the $\chi^2$ probability of the vertex fit, which 
is estimated by comparing results with and without this requirement. 
The results are summarized as detector effects in Table~\ref{sys1}. 
\begin{figure}[b]
\includegraphics[width=0.50\textwidth]{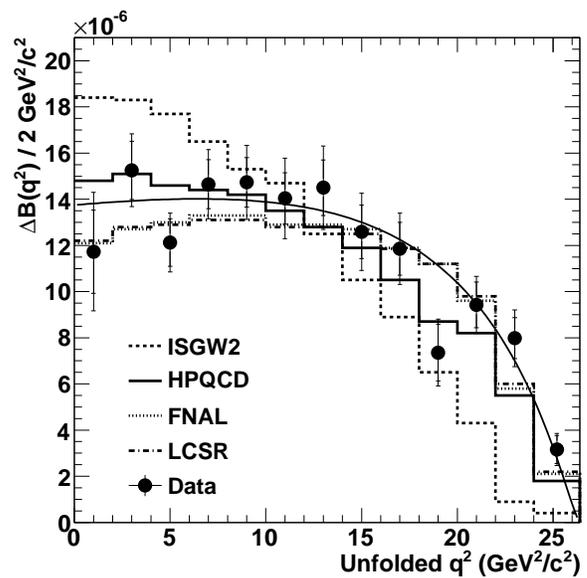}
\vspace{-9mm}
\caption{Distribution of the partial branching fraction as a function 
  of $q^2$ after unfolding~(closed circles). 
  The error bars show the statistical and the total
  uncertainty on the data. The curve is the result of a fit of the BK form
  factor parameterization~\cite{Becirevic:1999kt} to our data. The four 
  histograms
  (dashed:ISGW2; plain:HPQCD; dotted:FNAL; dot-dashed:LCSR) 
  show various form factor predictions.} \label{fig2}
\end{figure}
They depend weakly on $q^2$ and amount to 3.4\% for
the entire $q^2$~range. We vary the branching fractions of the
decays contributing to the $b\to u\ell\nu$ and $b\to
c\ell\nu$~backgrounds within $\pm 1$ standard deviation of their 
world-average values~\cite{Amsler:2008zzb} and assign an uncertainty of 0.6\%
to the total yield. We further consider form factor uncertainties in the
decays $B^0\to\pi^-\ell^+\nu$~\cite{Aubert:2005cd},
~$B^0\to\rho^-\ell^+\nu$~\cite{Ball:2004ye,Ball:2004rg}, 
$B^0\to D^-\ell^+\nu$ 
and $B^0\to D^{*-}\ell^+\nu$~\cite{TheHeavyFlavorAveragingGroup:2010qj}, 
and uncertainties in the shape function parameters of the inclusive 
$b\to u\ell\nu$~model~\cite{De Fazio:1999sv}. 
These uncertainties correspond to a 1.1\% error on $\mathcal{B}(B^0\to\pi^-\ell^+\nu)$. 
The uncertainty in the correction of the continuum MC is estimated by varying 
its weights by their statistical uncertainties. The other sources of systematic
uncertainty in Table~\ref{sys1} include the uncertainty 
in the $\Upsilon(4S)\to B^0\bar B^0$ branching fraction~\cite{Amsler:2008zzb}, 
limited MC statistics, the effect of
final state radiation, which is estimated by investigating MC samples with 
and without bremsstrahlung corrections calculated using the PHOTOS package, 
and the uncertainty in the number of $B\bar B$~pairs in the data sample. 
For values of $\Delta\mathcal{B}$ in individual $q^2$~bins, a breakdown of the
systematic uncertainties is presented in Table~\ref{breakdown.t} 
and the statistical and systematic correlations is given 
in Table~\ref{corr.stat.t} and Table~\ref{corr.syst.t}.

We fit the $\Delta \mathcal{B}$ distribution using the two-parameter
BK parameterization~\cite{Becirevic:1999kt} of $f_+(q^2)$, 
taking into account statistical and systematic correlations. 
The result is shown in Fig.~\ref{fig2}. 
Although this parameterization has been criticized~\cite{Becher-Hill:2006}, 
we present the fit result in order to directly compare with other existing 
results~\cite{Aubert:2006px}. 
We obtain 
$|V_{ub}|f_+(0)=(9.24\pm 0.18{(\mathrm{stat})}\pm 0.21{(\mathrm{syst})})\times 10^{-4}$ 
and 
$\alpha= 0.60\pm 0.03{(\mathrm{stat})}\pm 0.02{(\mathrm{syst})}$, 
where $\alpha$ is a positive constant that scales with $m_B$~\cite{Becirevic:1999kt}. 
The
$\chi^2$~probability of the fit is 62\%. We also calculate the
$\chi^2$~probabilities of different
theoretical form factor predictions with our binned data. 
We obtain probabilities of 42\%
and 43\% for the HPQCD~\cite{Dalgic:2006dt} and the FNAL~\cite{Okamoto:2004xg}
lattice QCD calculations, respectively, and 49\% for the LCSR
theory~\cite{Ball:2004ye}. The ISGW2 quark model~\cite{Scora:1995ty}, 
for which the probability is 
2.3$\times10^{-6}$, 
is incompatible with the experimental data.

As described in Ref.~\cite{Bailey:2008wp}, the CKM matrix
element $|V_{ub}|$ can be extracted from a simultaneous fit to 
experimental and lattice QCD results 
(from the FNAL/MILC Collaboration~\cite{Bailey:2008wp}), 
taking into account statistical and systematic correlations. 
To this end, the $q^2$ variable is transformed to a dimensionless variable 
$z$~\cite{Becher-Hill:2006,Boyd:1994tt;Boyd-Savage:1997}. 
In addition, the two functions, $P_+$ and $\phi_+$ are taken from 
Ref.~\cite{Arnesen:2005}, 
where $P_+$ is a function that accounts for the pole at $q^2=m_{B^*}^2$ 
and $\phi_+$ is an analytic function that controls the values of 
the $a_i$ series coefficients. 
In terms of the new variable $z$, the product of the form factor $f_+(q^2)$ 
and the functions $P_+$ and $\phi_+$ has the simple form, 
$\sum_{i=0}^{\infty}a_iz^i$. 
\begin{figure}[b]
\includegraphics[width=0.50\textwidth]{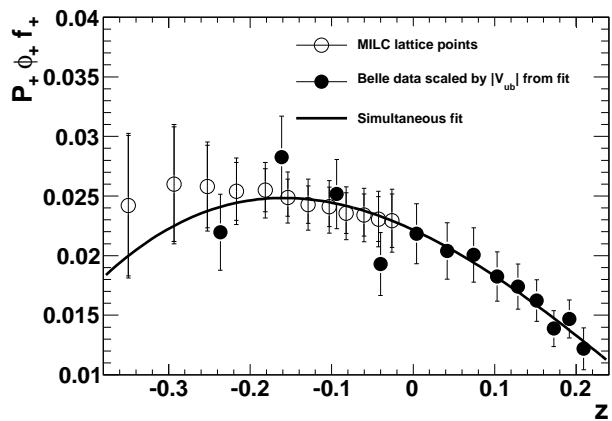}
\vspace{-7mm}
\caption{$|V_{ub}|$ extraction from a simultaneous fit of experimental~(closed circles)
  and FNAL/MILC lattice QCD results~(open circles)~\cite{Bailey:2008wp}. 
  The error for each experimental data point is the total experimental uncertainty. 
	The smaller error bars of the lattice QCD results are statistical only 
  while the larger ones also include systematic uncertainties.} 
\label{p-phi-fp-belle-milc.fig}
\end{figure}
We fit the lattice QCD results and experimental data with a third-order 
polynomial where the free parameters of the fit are the coefficients $a_i$ 
and the relative normalization between lattice QCD results 
and experimental results, which is $|V_{ub}|$. 
The resulting experimental data (which are scaled by the fitted $|V_{ub}|$ value) 
and the lattice QCD results 
are shown in Fig.~\ref{p-phi-fp-belle-milc.fig}. 
We obtain
$|V_{ub}|=(3.43\pm 0.33)\times 10^{-3}$,
$a_0 = 0.022 \pm 0.002$, 
$a_1 =-0.032 \pm 0.004$, 
$a_2 =-0.080 \pm 0.020$ 
and $a_3 = 0.081 \pm 0.066$, 
where the $\chi^2/\mathrm{n.d.f.}$ of the fit is approximately $12 / 20$. 
Statistically, we find no significant difference in the fitted value of $|V_{ub}|$ 
using second- and fourth-order polynomial fits. 
Note that the error in $|V_{ub}|$ includes both experimental and theoretical uncertainties. 
We find that the error includes a 3\% contribution from the branching fraction measurement, 
a 4\% from $q^2$ shape measured in data, 
and an 8\% uncertainty from theoretical normalization. 
The experimental and the total errors are compatible with the previous 
results in Ref.~\cite{Bailey:2008wp,Aubert:2010px}.


Alternatively, $|V_{ub}|$ can be determined from the measured partial branching fraction 
using the relation
$|V_{ub}|=\sqrt{\Delta\mathcal{B}/(\tau_{B^0}\Delta \zeta )}$, where
$\tau_{B^0}$ is the $B^0$~lifetime~\cite{Amsler:2008zzb} and
$\Delta\zeta$ is the normalized partial decay width derived in
different theoretical
approaches~\cite{Dalgic:2006dt,Okamoto:2004xg,Ball:2004ye}. These
calculations typically assume a specific parameterization of the form
factor shape.~Values of $|V_{ub}|$ for different form factor predictions 
are given in Table~\ref{ff-vub}.
\renewcommand{\arraystretch}{1.3}
\begin{table}[t]
\caption{Values extracted for $|V_{ub}|$ using different form factor predictions. 
  The first error on $|V_{ub}|$ is the experimental error including 
  statistical, systematic uncertainties and the uncertainty in the $B^0$~
  lifetime~\cite{Amsler:2008zzb}, the last asymmetric errors arise from the 
  uncertainty in $\Delta\zeta$.}
\begin{ruledtabular}
\begin{tabular}
{rccc}
$f_+(q^2)$                 & $q^2$ (GeV$^2/c^2$) & $\Delta\zeta$ (ps$^{-1}$) & $|V_{ub}|$
(10$^{-3}$) \\
\hline
HPQCD~\cite{Dalgic:2006dt} &  $>16$  & $2.07\pm 0.57$  & $3.55\pm 0.13$$^{+0.62}_{-0.41}$ \\
FNAL~\cite{Okamoto:2004xg} &  $>16$  & $1.83\pm 0.50$  & $3.78\pm 0.14$$^{+0.65}_{-0.43}$ \\
LCSR~\cite{Ball:2004ye}    &  $<16$  & $5.44\pm 1.43$  & $3.64\pm 0.11$$^{+0.60}_{-0.40}$ \\
\end{tabular}
\end{ruledtabular}
\label{ff-vub}
\end{table}

In summary, using 657$\times 10^6$ $B\bar{B}$ events of Belle $\Upsilon(4S)$~data 
we measure the partial branching fractions of the decay
$B^0\to\pi^-\ell^+\nu$ in 13 bins of $q^2$. The total branching fraction is found to be 
$(1.49\pm 0.04{(\mathrm{stat})}\pm0.07{(\mathrm{syst})})\times 10^{-4}$. 
A combined fit of experimental and
FNAL/MILC lattice QCD results~\cite{Bailey:2008wp}, yields a new precise
determination of $|V_{ub}|$ from this decay, $|V_{ub}|=(3.43\pm 0.33)\times 10^{-3}$. 
Determinations using only a fraction of the
phase space lead to less precise but statistically compatible numbers 
for $|V_{ub}|$: 
using a LCSR calculation for the region $q^2<16$~GeV$^2/c^2$~\cite{Ball:2004ye} yields 
$(3.64\pm 0.06{(\mathrm{stat})}\pm 0.09{(\mathrm{syst})}$$^{+0.60}_{-0.40}(\mathrm{FF}))\times 10^{-3}$. 
Assuming the HPQCD~\cite{Okamoto:2004xg} and the FNAL~\cite{Dalgic:2006dt} 
lattice QCD calculations,
sensitive to the region $q^2>16$~GeV$^2/c^2$, we obtain 
$(3.55\pm 0.09{(\mathrm{stat})}\pm 0.09{(\mathrm{syst})}$$^{+0.62}_{-0.41}(\mathrm{FF}))\times 10^{-3}$ 
and
$(3.78\pm 0.10{(\mathrm{stat})}\pm 0.10{(\mathrm{syst})}$$^{+0.65}_{-0.43}(\mathrm{FF}))\times 10^{-3}$, 
respectively.


We thank the KEKB group for excellent operation of the
accelerator, the KEK cryogenics group for efficient solenoid
operations, and the KEK computer group and
the NII for valuable computing and SINET3 network support.  
We acknowledge support from MEXT, JSPS and Nagoya's TLPRC (Japan);
ARC and DIISR (Australia); NSFC (China); MSMT (Czechia);
DST (India); MEST, NRF, NSDC of KISTI, and WCU (Korea); MNiSW (Poland); 
MES and RFAAE (Russia); ARRS (Slovenia); SNSF (Switzerland); 
NSC and MOE (Taiwan); and DOE (USA).
E.~Won acknowledges support by NRF Grant No. 2009-0071072.

\begin{table*}[hbtp]
\begin{center}
\caption{Correlation coefficients of the
  $\Delta\mathcal{B}(B^0\to\pi^-\ell^+\nu)$ statistical errors.}
\begin{ruledtabular}
\begin{tabular}{c|rrrrrrrrrrrrr}
$q^2$(GeV$^2/c^2$) & 0-2   & 2-4  & 4-6   & 6-8   & 8-10  & 10-12 & 12-14 & 14-16 & 16-18 & 18-20 & 20-22 & 22-24 & 24-26.4 \\
\hline
    0-2 &  1.000   & -0.275   &  0.109   &  0.036   &  0.021   &  0.020   &  0.031   & -0.005   & -0.001   & -0.002   & -0.003   & -0.004   & -0.007 \\
    2-4 & -0.275   &  1.000   & -0.292   &  0.168   & -0.008   &  0.026   &  0.029   & -0.005   & -0.001   & -0.002   & -0.004   & -0.003   & -0.004 \\
    4-6 &  0.109   & -0.292   &  1.000   & -0.229   &  0.156   &  0.049   &  0.101   & -0.014   & -0.003   & -0.005   & -0.004   & -0.004   & -0.002 \\
    6-8 &  0.036   &  0.168   & -0.229   &  1.000   & -0.255   &  0.123   &  0.064   & -0.009   & -0.003   & -0.004   & -0.003   & -0.002   & -0.003 \\
   8-10 &  0.021   & -0.008   &  0.156   & -0.255   &  1.000   & -0.156   &  0.225   & -0.032   & -0.001   & -0.004   & -0.004   & -0.004   & -0.006 \\
  10-12 &  0.020   &  0.026   &  0.049   &  0.123   & -0.156   &  1.000   & -0.054   &  0.023   & -0.007   & -0.002   & -0.006   & -0.007   & -0.009 \\
  12-14 &  0.031   &  0.029   &  0.101   &  0.064   &  0.225   & -0.054   &  1.000   & -0.239   &  0.007   & -0.028   & -0.007   & -0.011   & -0.013 \\
  14-16 & -0.005   & -0.005   & -0.014   & -0.009   & -0.032   &  0.023   & -0.239   &  1.000   & -0.013   &  0.144   & -0.030   & -0.006   & -0.010 \\
  16-18 & -0.001   & -0.001   & -0.003   & -0.003   & -0.001   & -0.007   &  0.007   & -0.013   &  1.000   &  0.101   & -0.028   & -0.002   & -0.004 \\
  18-20 & -0.002   & -0.002   & -0.005   & -0.004   & -0.004   & -0.002   & -0.028   &  0.144   &  0.101   &  1.000   & -0.167   & -0.025   & -0.040 \\
  20-22 & -0.003   & -0.004   & -0.004   & -0.003   & -0.004   & -0.006   & -0.007   & -0.030   & -0.028   & -0.167   &  1.000   & -0.036   &  0.006 \\
  22-24 & -0.004   & -0.003   & -0.004   & -0.002   & -0.004   & -0.007   & -0.011   & -0.006   & -0.002   & -0.025   & -0.036   &  1.000   & -0.111 \\
24-26.4 & -0.007   & -0.004   & -0.002   & -0.003   & -0.006   & -0.009   & -0.013   & -0.010   & -0.004   & -0.040   &  0.006   & -0.111   &  1.000 \\
\end{tabular}
\end{ruledtabular}
\label{corr.stat.t}
\end{center}
\end{table*}

\begin{table*}[hbt]
\begin{center}
\caption{Correlation coefficients of the
  $\Delta\mathcal{B}(B^0\to\pi^-\ell^+\nu)$ systematic errors.}
\begin{ruledtabular}
\begin{tabular}{c|rrrrrrrrrrrrr}
$q^2$(GeV$^2/c^2$) & 0-2     & 2-4     &  4-6    & 6-8     & 8-10    & 10-12   & 12-14   & 14-16   & 16-18   & 18-20   & 20-22   & 22-24   & 24-26.4    \\
\hline
   0-2  &  1.000  & -0.256  &  0.187  & -0.162  &  0.297  &  0.181  &  0.224  &  0.114  &  0.104  &  0.112  &  0.084  &  0.020  &  0.069 \\
   2-4  & -0.256  &  1.000  &  0.142  &  0.570  &  0.075  &  0.163  &  0.162  &  0.193  &  0.202  &  0.210  &  0.244  &  0.303  &  0.282 \\
   4-6  &  0.187  &  0.142  &  1.000  &  0.202  &  0.459  &  0.451  &  0.469  &  0.212  &  0.368  &  0.329  &  0.332  &  0.322  &  0.336 \\
   6-8  & -0.162  &  0.570  &  0.202  &  1.000  & -0.017  &  0.240  &  0.202  &  0.280  &  0.256  &  0.284  &  0.312  &  0.333  &  0.272 \\
  8-10  &  0.297  &  0.075  &  0.459  & -0.017  &  1.000  &  0.375  &  0.633  &  0.156  &  0.321  &  0.290  &  0.258  &  0.244  &  0.231 \\
 10-12  &  0.181  &  0.163  &  0.451  &  0.240  &  0.375  &  1.000  &  0.433  &  0.214  &  0.328  &  0.332  &  0.252  &  0.284  &  0.230 \\
 12-14  &  0.224  &  0.162  &  0.469  &  0.202  &  0.633  &  0.433  &  1.000  & -0.013  &  0.337  &  0.278  &  0.291  &  0.245  &  0.246 \\
 14-16  &  0.114  &  0.193  &  0.212  &  0.280  &  0.156  &  0.214  & -0.013  &  1.000  &  0.334  &  0.500  &  0.287  &  0.337  &  0.322 \\
 16-18  &  0.104  &  0.202  &  0.368  &  0.256  &  0.321  &  0.328  &  0.337  &  0.334  &  1.000  &  0.452  &  0.334  &  0.356  &  0.344 \\
 18-20  &  0.112  &  0.210  &  0.329  &  0.284  &  0.290  &  0.332  &  0.278  &  0.500  &  0.452  &  1.000  &  0.208  &  0.430  &  0.333 \\
 20-22  &  0.084  &  0.244  &  0.332  &  0.312  &  0.258  &  0.252  &  0.291  &  0.287  &  0.334  &  0.208  &  1.000  &  0.222  &  0.402 \\
 22-24  &  0.020  &  0.303  &  0.322  &  0.333  &  0.244  &  0.284  &  0.245  &  0.337  &  0.356  &  0.430  &  0.222  &  1.000  &  0.220 \\
24-26.4 &  0.069  &  0.282  &  0.336  &  0.272  &  0.231  &  0.230  &  0.246  &  0.322  &  0.344  &  0.333  &  0.402  &  0.220  &  1.000 \\
\end{tabular}
\end{ruledtabular}
\label{corr.syst.t}
\end{center}
\end{table*}

\begin{sidewaystable}[p]
\begin{center}
\hspace{15mm}
\vspace{8cm}
\caption{The raw, unfolded yields, signal efficiencies and the partial branching fractions and their relative errors
  (\%) from various sources in 13 bins of $q^2$.}
\begin{ruledtabular}
\begin{tabular}{lcccccccccccccccc}
$q^2$(GeV$^2/c^2$) & 0-2   & 2-4   & 4-6   & 6-8   & 8-10  & 10-12 & 12-14 & 14-16 & 16-18 & 18-20 & 20-22 & 22-24  & 24-26.4 & 0-16 & 16-26.4 &  Total\\
\hline
\hline

Raw yields      & 1225.80  & 1788.46 & 1756.48 & 2074.77 & 2100.64 & 2136.45 & 2139.00  & 1951.65 & 1537.95 & 1070.35 & 1422.80 & 1369.84 & 912.27 
                & 15173.25 & 6313.21 & 21486.46 \\
Unfolded yields & 1179.95  & 1873.57 & 1662.57 & 2141.65 & 2107.29 & 2192.33 & 2243.97 & 2066.70 & 1609.10 & 1011.85 & 1452.82 & 1323.35 & 621.30 
                & 15468.02 & 6018.44 & 21486.46 \\
Efficiencies (\%) &   7.66 &   9.35 &  10.44 &  11.13 &  10.89 &  11.89 &  11.78 &  12.50 &  10.33 &  10.47 &  11.74 &  12.63 &  14.98 & & \\
$\Delta\mathcal{B}$ ($\times$ 10$^7$) & 117.33 & 152.58 & 121.28 & 146.54 & 147.32 & 140.39 & 145.00 & 125.90 & 118.57 & 73.59 & 94.21 & 79.80 & 31.59   & 1096.34 &  397.75 & 1494.09\\
\hline
Lepton ID             & 2.35    & 2.41    & 2.39    & 2.41    & 2.38    & 2.38    & 2.43    & 2.47    & 2.47    & 2.49    & 2.45    & 2.44    & 2.56    & 2.40    & 2.49    & 2.44   \\
Pion ID               & 1.32    & 1.37    & 1.43    & 1.49    & 1.41    & 1.36    & 1.32    & 1.24    & 1.11    & 0.97    & 0.85    & 1.12    & 1.38    & 1.37    & 1.08    & 1.26   \\
Tracking efficiency   & 2.00    & 2.00    & 2.00    & 2.00    & 2.00    & 2.00    & 2.00    & 2.00    & 2.00    & 2.00    & 2.00    & 2.01    & 2.45    & 2.00    & 2.09    & 2.04    \\
$\gamma$ efficiency  & 0.21   & 0.41   & 0.14   & 0.25   & 0.32   & 0.45   & 0.94   & 0.24   & 0.81   & 0.80   & 0.23   & 0.26   & 0.49    & 0.37 & 0.51 & 0.42 \\
Vertex $\chi^2$ probability & 0.15   & 0.15   & 0.15   & 0.15   & 0.15   & 0.15   & 0.15   & 0.15   & 0.15   & 0.15   & 0.15   & 0.15   & 0.15    & 0.15 & 0.15 & 0.15 \\
\hline
$B \to \rho \ell \nu$ BF   & 0.58    & 0.60    & 0.59    & 0.46    & 0.74    & 0.60    & 0.41  & 0.57    & 0.48    & 0.47    & 0.41    & 0.41    & 0.33   &  0.44  &  0.42  & 0.43    \\
$B \to \omega \ell \nu$ BF & 0.28    & 0.16    & 0.14    & 0.12    & 0.13    & 0.11    & 0.16  & 0.08    & 0.09    & 0.12    & 0.30    & 0.52    & 1.19   &  0.11  &  0.31  & 0.16    \\
$B \to b_1 \ell \nu$ BF    & 0.30    & 0.16    & 0.14    & 0.12    & 0.13    & 0.12    & 0.13  & 0.09    & 0.11    & 0.12    & 0.11    & 0.14    & 0.59   &  0.14  &  0.14  & 0.14    \\
$V_{ub}$ + other $X_u \ell \nu$ BF & 4.43    & 1.55    & 0.96    & 0.87    & 0.45    & 0.35    & 0.45  & 0.61    & 0.13    & 0.13    & 0.12    & 0.17    & 0.77   &  0.19  &  0.15  & 0.15    \\
$B \to D^* \ell \nu$ BF     & 0.42    & 0.40    & 0.15    & 0.67    & 0.09    & 0.12    & 0.20   & 0.09    & 0.16    & 0.12    & 0.12    & 0.14    & 0.27  &  0.18  &  0.13  & 0.16    \\
$B \to D \ell \nu$ BF       & 0.27    & 0.14    & 0.14    & 0.12    & 0.12    & 0.09    & 0.07   & 0.12    & 0.09    & 0.14    & 0.14    & 0.18    & 0.61  &  0.07 &   0.14  & 0.08    \\
$B \to D^{**} \ell \nu$ BF  & 0.20    & 0.16    & 0.16    & 0.14    & 0.14    & 0.12    & 0.11   & 0.11    & 0.42    & 0.15    & 0.14    & 0.20    & 0.11  &  0.11  &  0.22  & 0.13    \\
Other $X_c \ell \nu$ BF        & 0.13    & 0.09    & 0.13    & 0.09    & 0.18    & 0.12    & 0.13   & 0.09    & 0.14    & 0.14    & 0.14    & 0.20    & 0.12  &  0.06  &  0.13  & 0.06    \\
\hline
$B^0\to\pi^- \ell^+ \nu$ FF  & 3.58    & 1.64    & 1.26    & 1.27    & 1.44    & 1.57    & 1.67    & 1.70    & 1.78    & 1.97    & 1.61    & 1.92    & 4.03    & 0.63    & 0.86    & 0.53    \\
$B^0\to\rho^- \ell^+ \nu$ FF & 3.59    & 1.73    & 1.51    & 1.47    & 1.64    & 1.82    & 2.04    & 1.89    & 2.01    & 2.30    & 1.99    & 2.50    & 4.98    & 0.72    & 0.95    & 0.60    \\
SF parameter                              & 1.44    & 0.63    & 1.59    & 1.07    & 2.10    & 2.80    & 2.52    & 2.42    & 2.24    & 4.02    & 2.12    & 3.15    & 4.66    & 0.71    & 1.17    & 0.63    \\
$B^0\to D^{*-} \ell^+ \nu$ FF   & 0.51    & 0.64    & 0.60    & 0.89    & 1.54    & 1.77    & 2.51   & 0.81    & 1.18    & 0.56    & 0.51    & 0.42    & 0.98   &  0.48   &  0.34   & 0.36    \\
$B^0\to D^- \ell^+ \nu$ FF     & 0.33    & 0.09    & 0.20    & 0.30    & 0.24    & 0.12    & 0.26   & 0.20    & 0.20    & 0.09    & 0.16    & 0.11    & 0.11   &  0.10   &  0.08   & 0.08    \\
\hline
$\Upsilon(4S)\to B^0\bar B^0$ BF & 2.11    & 1.39    & 2.41    & 3.28    & 3.68    & 3.90    & 3.67    & 2.76    & 2.83    & 4.58    & 3.62    & 5.21    & 2.28    & 1.56    & 1.72    & 1.40    \\
Signal MC stat. error                      & 0.48    & 0.15    & 0.23    & 0.27    & 0.30    & 0.24    & 0.28    & 0.28    & 0.38    & 0.52    & 0.58    & 0.74    & 1.97    & 0.12    & 0.39    & 0.15    \\

\hline
FSR               & 0.31    & 0.58    & 1.03    & 0.88    & 0.93    & 1.24    & 1.18    & 1.43    & 1.01    & 1.23    & 1.15    & 0.78    & 0.66   &  0.45  &  0.56   & 0.37    \\
$B$ counting        & 1.36    & 1.36    & 1.36    & 1.36    & 1.36    & 1.36    & 1.36    & 1.36    & 1.36    & 1.36    & 1.36    & 1.36    & 1.36    & 1.36    & 1.36    & 1.36    \\

\hline
Continuum $q^2$       &13.33    & 3.64    & 3.01    & 4.35    & 4.91    & 6.62    & 6.04 & 7.42    & 6.22    & 6.60    & 4.67    & 7.13    & 5.90    & 2.14  & 2.62  & 1.80    \\
\hline
\hline
Total systematic error & 15.63   & 6.23   & 6.14   & 7.19   & 8.04   & 9.56   & 9.27   &  9.60   & 8.71   & 10.30   & 7.80   & 10.63   & 11.23  & 4.78 & 5.26 & 4.54 \\
Total statistical error  & 15.35   & 8.27   & 8.50   & 7.24   & 7.27   & 7.91   & 8.32   & 9.19   &  9.67   & 16.68   & 10.54   & 11.12   & 18.98 & 3.03 & 5.31 & 2.63 \\
Total error            & 21.90   & 10.35   & 10.48   & 10.20   & 10.84   & 12.41   & 12.46   & 13.29   & 13.01   & 19.60   & 13.12   & 15.39   & 22.08 & 5.66 & 7.47 & 5.24 \\
\end{tabular}
\end{ruledtabular}
\label{breakdown.t}
\end{center}
\end{sidewaystable}



\begin{thebibliography}{99}
\bibitem{CC}
  Throughout this paper, the inclusion of the charge-conjugate decay mode
  is implied unless otherwise stated.
\end{thebibliography}
\end{document}